\documentclass[conference]{IEEEtran}
\IEEEoverridecommandlockouts

\usepackage{standalone}
\usepackage[reqno]{amsmath}
\usepackage{amssymb}

\usepackage{algorithm}
\usepackage{algorithmicx}
\usepackage{algpseudocode}
\usepackage{setspace}
\usepackage{multicol}
\usepackage{enumerate}
\usepackage[font=small, belowskip=-15pt,aboveskip=0pt]{caption}
\usepackage{cite}
\usepackage{amsthm}
\usepackage[]{graphicx}
\usepackage{tikz}
\usepackage{tkz-tab}
\usepackage{pgfplots}
\usepackage{booktabs}
\usepackage{floatflt}

\usepackage{psfrag}	
\usepackage{array}
\usepackage{multirow,hhline}
\usepackage{exscale}
\usepackage{color}
\usepackage{colortbl}
\usepackage{epsfig}

\usepackage{array}
\usepackage[tight]{subfigure}

\usepackage[latin1]{inputenc} 
\usepackage[T1]{fontenc} 
\usepackage{rotating}
\usepackage{pbox}

\usepackage{pifont}

\usepackage{comment}
\usepackage{bookmark}
\usepackage{cleveref}

\newcommand{\mat}[1]{\ensuremath{\boldsymbol{#1}}}
\renewcommand{\vec}[1]{\ensuremath{\boldsymbol{#1}}}

\DeclareMathOperator{\diag}{diag}

\usetikzlibrary{shapes,snakes}
\usetikzlibrary{calc}
\usetikzlibrary{patterns}
\usetikzlibrary{decorations.pathmorphing} 
\usetikzlibrary{spy}
\usetikzlibrary{positioning}
\usetikzlibrary{arrows, decorations.markings}
\usetikzlibrary{shapes.multipart, shapes.geometric, fit, backgrounds}
\usetikzlibrary{spy}

\definecolor{mycolor1}{rgb}{0.00000,0.44700,0.74100}%
\definecolor{mycolor2}{rgb}{0.85000,0.32500,0.09800}%
\definecolor{mycolor4}{rgb}{0.92900,0.69400,0.12500}%
\definecolor{mycolor3}{rgb}{0.49400,0.18400,0.55600}%
\definecolor{mycolor5}{rgb}{0.46600,0.67400,0.18800}%
\definecolor{mycolor6}{rgb}{0.30100,0.74500,0.93300}%
\definecolor{mycolor7}{rgb}{0.63500,0.07800,0.18400}%

\begin{document}

\title{Path Loss Compensation vs. Spatial Diversity: Mixed-Criticality Superpostion Coding in RIS-assisted THz communication}
\title{Exploiting Multipath Diversity in sub-THz Communication: A Mixed-Criticality Superposition Coding Scheme}
\title{Intermittency versus Path Loss in RIS-aided THz Communication: A Data Significance Approach}

\author{\IEEEauthorblockN{Yasemin Karacora, Adam Umra and Aydin Sezgin}
    \IEEEauthorblockA{Institute of Digital Communication Systems, Ruhr University Bochum, Germany \\ Emails: \{yasemin.karacora, adam.umra, aydin.sezgin\}@rub.de \vspace{-.5cm}}
    \thanks{This project has received funding from the programme ``Netzwerke 2021'', an initiative of the Ministry of Culture and Science of the State of Northrhine-Westphalia. The sole responsibility for the content of this publication lies with
the authors.}}

\maketitle

 \begin{abstract}
The transition to Terahertz (THz) frequencies, providing an ultra-wide bandwidth, is a key driver for future wireless communication networks. However, the specific properties of the THz channel, such as severe path loss and vulnerability to blockage, pose a significant challenge in balancing data rate and reliability. This work considers reconfigurable intelligent surface (RIS)-aided THz communication, where the effective exploitation of a strong, but intermittent line-of-sight (LOS) path versus a reliable, yet weaker RIS-path is studied. We introduce a mixed-criticality superposition coding scheme that addresses this tradeoff from a data significance perspective. The results show that the proposed scheme enables reliable transmission for a portion of high-criticality data without significantly impacting the overall achievable sum rate and queuing delay. Additionally, we gain insights into how the LOS blockage probability and the channel gain of the RIS-link influence the rate performance of our scheme. \looseness-1
\end{abstract}

\section{Introduction}
In order to meet the ultra-high data rate demands of future 6G applications, e.g., extended reality or digital twins, the migration to higher frequency bands, such as the Terahertz (THz) regime, becomes a promising solution. 
However, the THz band faces challenges like severe path loss, molecular absorption, high penetration and reflection losses, creating a highly uncertain and dynamic channel \cite{saad2021sevenTHz, akyildiz2018combating}. To overcome this key challenge, employing narrow pencil beams to focus the transmit power toward the receiver becomes necessary. 
Furthermore, the THz channel is primarily determined by the line-of-sight (LOS) path, whereas reflection paths are typically severely attenuated \cite{kokkoniemi2019NLOS300}. 
Yet, relying solely on the LOS path in order to achieve sufficient signal strength at the receiver may result in intermittent connectivity, as THz links are prone to blockage caused by moving obstacles or even the user itself. In addition, pencil-shaped beams entail a high risk for beam misalignment even due to micro-mobility of the user. However, in order to provide a seamless user experience, a high reliability as well as low latencies are essential for a majority of 6G applications, especially mission- and safety-critical use cases, such as healthcare applications or industrial automation.
Although these THz effects can partly be mitigated via robust beamforming schemes (e.g., \cite{karacoraTHzTCOM}) or the use of reconfigurable intelligent surfaces (RIS) (\cite{aman2023downlink, chaccour2020risk, zarini2023RISaided}) to overcome blockages, tackling the rate-reliability tradeoff in high frequency bands is still a fundamental challenge. 
Hence, this work aims to address the effective use of multipath components, particularly in RIS-aided systems, in a LOS-dominated THz channel. RIS-assisted channels provide the opportunity of establishing an alternative transmission link that is less affected by blockages due to appropriate placement of the RIS. Yet, despite beam steering at the RIS, the channel gain of such reflection paths is still significantly weaker than a direct LOS path. Hence, we are facing a tradeoff between utilizing the strong, but highly intermittent direct path, versus the weaker, yet more reliable RIS-path. 
While such a tradeoff is typically addressed by the attempt of finding a compromise between the different objectives, i.e., data rate and reliability, the paradigm shift in 6G towards goal-oriented communication schemes offers a new approach, namely leveraging service differentiation and prioritization techniques. That is, by taking account of the semantic meaning of the transmit data in the context of a specific task or goal, many applications typically comprise heterogeneous data in terms of their criticality and significance. Thus, an efficient way of tackling quality of service (QoS) tradeoffs is to establish criticality-aware transmission schemes that can serve the throughput required by an application, yet guarantees strictly reliable delivery for critical data segments. In this work, we consider the simultaneous arrival of data streams of high and low criticality, which are then processed in parallel, thereby enabling a differentiated approach to the rate-reliability tradeoff in RIS-assisted THz communications.

\subsection{Related Work}
The severe THz path loss and the susceptibility to blockage intensify the rate-reliability tradeoff in THz communication systems. This key challenge has been addressed in several works, e.g., \cite{chaccour2020HRLLC, boulogeorgos2021directional, chaccour2020risk}. The deployment of RIS has emerged as a promising mitigation strategy for the intermittent nature of THz links. For instance, RIS-assisted THz systems have been studied in \cite{aman2023downlink, chaccour2020risk, zarini2023RISaided}.
In \cite{chaccour2020risk}, RIS are deployed as THz base stations to enable a seamless user experience in virtual reality applications. The authors in \cite{zarini2023RISaided} consider the resource management in RIS-aided THz systems serving users with diverse QoS requirements. These works, however, do not study the intermittency and path loss in THz systems from a data significance perspective. 
Superpostion coding (SC) has been addressed in the context of THz communication, e.g., in \cite{sarieddeen2021terahertz}, where the transmission of multiple streams in a point-to-point link, as well as multi-user non-orthogonal multiple access (NOMA) has been studied. The application of SC for multi-resolution broadcasting has been proposed in \cite{choi2015multicast}. Different from \cite{choi2015multicast}, where two data streams of different priority are transmitted to two users in a multicast NOMA setting, our work is using SC to deliver mixed-criticality data to a single user while addressing the specific THz challenges, i.e., path loss and intermittency.
The need for developing criticality-aware communication schemes has been addressed in previous works, e.g., \cite{reifert2023comeback}, where the consideration of mixed-critical QoS levels has been shown to enhance resilience in resource management.
In \cite{karacora2023rate}, a criticality-aware superposition coding approach has been proposed to establish multi-connectivity in uplink rate-splitting for the purpose of ultra-reliable low-latency communication. \looseness-1

\subsection{Contribution}
In this paper, we examine the downlink of an RIS-assisted THz system, focusing on the rate-reliability tradeoff arising from LOS dominance in THz frequency bands and the severe intermittency of such LOS links. By proposing a mixed-criticality superposition coding (MC-SC) scheme, we address this tradeoff from a data significance perspective.

We consider a queuing model at the base station to capture the mixed-criticality data arrivals and formulate a power allocation optimization problem in order to stabilize the queues. 
The non-convex problem is solved iteratively by adopting a successive convex approximation method and a quadratic transform within a fractional programming framework \cite{shen2018fractional}. The performance is evaluated in terms of the maximum achievable data rate and the queuing delay and compared to an orthogonal baseline scheme. We demonstrate the potential of our proposed scheme, showing that a portion of the transmit data can be delivered with high reliability without increasing the delay and with relatively low sacrifices in terms of the maximum achievable sum rate. We further gain insights into how the degree of LOS intermittency as well as the number of RIS elements impact our scheme.

\section{System Model}
\begin{figure*}
    \centering
    \includegraphics{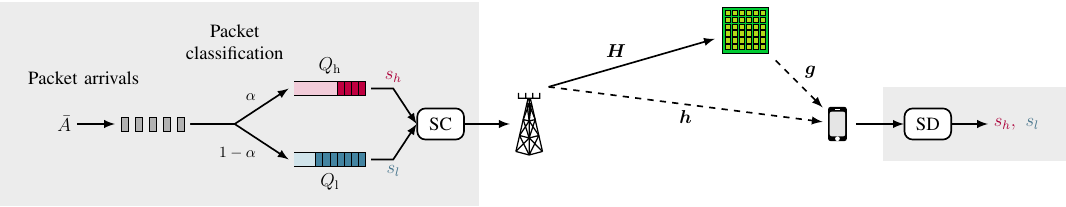}
    \vspace{-.5mm}
    \caption{Considered system model of a RIS-assisted BS-UE downlink channel. The arriving packets are classified according to their criticality and stored in a high- and a low-criticality (HC/LC) buffer. The BS applies superposition coding (SC) of the mixed-criticality data streams and the UE adopts a successive decoding (SD) approach.}
    \label{fig:sys_model}
\end{figure*}
We consider RIS-assisted THz downlink transmission, where a single base station (BS) transmits mixed-criticality data to a single user equipment (UE) via a direct LOS link and a reflective path provided by a RIS as depicted in Figure \ref{fig:sys_model}. The BS is equipped with $N_\mathrm{B}$ antennas with directivity gain $G_\mathrm{B}$, the UE has a single antenna with gain $G_\mathrm{U}$, and the RIS comprises $N_\mathrm{R}$ reflective elements.
At the BS, the data packets intended for the considered user are classified and assigned to one of two distinct criticality levels, denoted as high-criticality (HC) and low-criticality (LC). These may correspond to different applications or services, which have heterogeneous reliability and latency requirements. Alternatively, within a goal-oriented communication framework, data packets from a singular service can be categorized based on their significance. For instance, in a virtual reality (VR) application, data treated as HC may include real-time positional tracking and interactive feedback, which is essential for an immediate and responsive user experience. Less time-sensitive elements, like non-real-time rendering of static components, are categorized as LC, allowing for slightly delayed processing without compromising the overall VR experience.
Here, we define $\alpha \in [0,1]$ as the fraction of data classified as HC. 

\subsection{Channel Model}
We adopt the Saleh-Valenzuela channel model that is widely used for THz communications and generally consists of one LOS path and a few reflection paths \cite{sarieddeen2021overview}. However, due to the severe attenuation induced by scattering in the THz band, we neglect the non-line-of-sight (NLOS) component and only consider the dominant LOS path and one RIS-aided reflection path in our channel model.
Due to the high penetration loss THz channels are vulnerable to dynamic blockage caused by other objects or the user itself.  
Therefore, THz channels suffer from a significant random intermittency that can lead to frequent outages. We assume that for an appropriate RIS placement, the BS-RIS link is not affected by intermittency. Furthermore, the RIS is located much closer to the user than the BS. Thus, we assume that the path via the RIS is more reliable and only blocked when the direct path is obstructed.
We model the intermittency by a Bernoulli random variable $\beta_d$ ($\beta_r$), which is equal to one if the direct LOS path (RIS-path) between AP and UE is available, and equal to zero if the link is blocked. 
We define blockage probabilities $q_d = \text{Prob}(\beta_d=0)$ and $q_r= \text{Prob}(\beta_r=0)$, with $q_r < q_d$.
Hence, the direct link of the BS-UE channel is given by
\begin{equation}
    \vec{h} = \beta_d \eta_d \vec{a}_\mathrm{N_\mathrm{B}} (\varphi_\mathrm{BU}). 
\end{equation}
Here, $\eta_d$ and $\varphi_\mathrm{BU}$ are the path loss and the angle of departure (AoD) of the direct link, respectively, and $\vec{a}_{N} (\cdot)$ denotes the array response vector defined as $\vec{a}_N(\varphi) = [1, e^{j\pi \sin(\varphi)}, \dots, e^{j\pi(N-1)\sin(\varphi)}]^T$.
The channel gain of the direct THz link, comprising free space path loss and molecular absorption, is given as \cite{aman2023downlink}
\begin{equation}
    \mathrm{\eta}_d = \frac{\sqrt{G_\mathrm{B} G_\mathrm{U}} c}{4\pi f d_\mathrm{BU}} e^{-\frac{1}{2} k_a(f) d_\mathrm{BU}}, 
\end{equation}
where $f$ is the operating frequency, $c$ is the speed of light, and $d_\mathrm{BU}$ represents the distance between the BS and the UE. The frequency-dependent molecular absorption coefficient is denoted by $k_a(f)$ and is obtained based on the model given in \cite{kokkoniemi2021line}.
The BS-RIS-UE channel is obtained as
\begin{equation}
    \vec{g}^H \mat{\Phi} \mat{H} = \beta_r \eta_r \vec{a}_\mathrm{N_\mathrm{R}}^H(\varphi_\mathrm{RU}) \mat{\Phi} \vec{a}_\mathrm{N_\mathrm{R}}^H(\varphi_\mathrm{RB}) \vec{a}_\mathrm{N_\mathrm{B}}^H(\varphi_\mathrm{BR}),
\end{equation}
in which $\mat{H} \in \mathbb{C}^{N_\mathrm{B} \times N_\mathrm{R}}$ and $\vec{g} \in \mathbb{C}^{N_\mathrm{R} \times 1}$ represent the channel of the BS-RIS link and the RIS-UE link, respectively, while $\mat{\Phi} = \diag([e^{j\phi_1},\dots,e^{j\phi_{N_\mathrm{R}}}])$ is the phase shift matrix of the RIS. The AoD of the RIS-UE path, and the AoD and the angle of arrival of the BS-RIS path, are defined as $\varphi_{RU}$, $\varphi_{BR}$ and $\varphi_{RB}$, respectively. 
Here, the channel gain of the RIS-path is obtained as \cite{aman2023downlink}
\begin{equation}
     \mathrm{\eta}_r = \frac{\sqrt{G_\mathrm{B} G_\mathrm{U}} L_x L_y}{4\pi d_\mathrm{BR} d_\mathrm{RU} } e^{-\frac{1}{2} k_a(f) (d_\mathrm{BR}+d_\mathrm{RU})},
\end{equation}
in which $d_\mathrm{BR}$ and $d_\mathrm{RU}$ denote the BS-RIS and RIS-UE distances, while $L_x$ and $L_y$ are the length and width of a single RIS element, respectively, and assumed to be half of a wavelength. 

Hence, the received signal at the UE can be written as
\begin{equation}
    y = \left(\vec{h}^H + \vec{g}^H \mat{\Phi} \mat{H}\right) \vec{x} + n.
\end{equation}
Here, $\vec{x} \in \mathbb{C}^{N_\mathrm{B} \times 1}$ denotes the transmit signal and $n \sim \mathcal{CN}(0, \sigma_n^2)$ is additive white Gaussian noise (AWGN).

\subsection{Queuing Model}
As shown in Figure \ref{fig:sys_model}, we apply a queuing model at the BS, where packet arrivals are assumed to follow a Poisson process with arrival rate $\Bar{A}$. The arriving packets are then classified according to their criticality level, whereby a fraction of $\alpha$ data packets is considered HC data. We model the evolution of the two data queues containing HC and LC packets as follows:
\begin{align}
    Q_\mathrm{h}(t) &= \left[Q_\mathrm{h}(t-1) - \frac{T}{M}R_\mathrm{h}(t) \right]^+ + \alpha A(t),\\
    Q_\mathrm{l}(t) &= \left[Q_\mathrm{l}(t-1) - \frac{T}{M}R_\mathrm{l}(t) \right]^+ + (1-\alpha) A(t).
\end{align}
Here, $Q_\mathrm{h}(t)$ and $Q_\mathrm{l}(t)$ are the number of buffered HC and LC packets at time slot $t$, and $A(t)$ represents the total packet arrivals in time slot $t$. The packet size is denoted by $M$, and $T$ is the time slot duration.

\subsection{Mixed-Criticality Superposition Coding Scheme}
In our proposed scheme, superposition coding is applied to the mixed-criticality data, which is then simultaneously transmitted using the same frequency resources. Here, significantly more power is allocated to the HC data, enabling it to be decoded in a weaker channel, while LC data is successively decoded only in a sufficiently strong channel. When applied to a RIS-aided THz channel with LOS intermittency, our scheme allows the decoding of the critical stream even if the direct path is blocked, whereas non-critical data is only decodable in case of LOS availability.
\footnote{Note that power domain NOMA uses the same underlying concept to jointly serve a strong (near) and a weak (far) user. While we only consider one user in our MC-SC scheme, the user acts as a "strong user" (decoding both messages) when the direct LOS path is available, and as a "weak user" (decoding the strong/critical message only) when the LOS-link is blocked.}
To this end, the HC and LC messages, denoted by $s_h$ and $s_l$, respectively, are encoded so that the transmit signal is obtained as
\begin{equation}
    \vec{x} = \vec{f}_h s_h + \vec{f}_l s_l,
\end{equation}
where $\vec{f}_h \in \mathbb{C}^{N_\mathrm{B}\times 1}$ and $\vec{f}_l \in \mathbb{C}^{N_\mathrm{B}\times 1}$ are the precoding vectors of the HC and LC message. 
The user successively decodes both messages, starting with the HC message.
Thus, the signal-to-interference-plus-noise ratio (SINR) of the HC message can be written as
\begin{equation}
\label{sinr_h}
    \Gamma_h = \frac{\left|\left(\vec{h} + \vec{g}^H\mat{\Phi}\mat{H}\right) \vec{f}_h\right|^2}{\left|\left( \vec{h} + \vec{g}^H\mat{\Phi}\mat{H}\right) \vec{f}_l\right|^2 + \sigma_n^2}.
\end{equation}
After correctly decoding and subtracting the HC message from the received signal, the LC data can be decoded, whereby the signal-to-noise-ratio (SNR) is given as
\begin{equation}
\label{sinr_l}
    \Gamma_l = \frac{\left|\left( \vec{h} + \vec{g}^H\mat{\Phi}\mat{H}\right) \vec{f}_l\right|^2}{\sigma_n^2}.
\end{equation}

For tractability, within the scope of this paper, we assume that perfect channel state information (CSI) is available at the BS in order to ideally select the transmit beamformers of the BS as well as the phase shifts of the RIS. Given the LOS-dominance of the THz channel and assuming that narrow pencil beams are utilized to overcome the severe path loss at such high frequencies, we consider perfectly aligned beams without any significant reflections or scattering effects.\footnote{Note that this work aims at generally examining the potential benefits of adopting a data significance approach to tackle THz-specific challenges. An analysis of the performance of our scheme under more practical assumptions, e.g., user mobility and imperfect beam alignment, is subject to future work.}
Based on these assumptions, we can simplify the SINR-expressions as explained in the following. 
Let 
\begin{align}
    \vec{f}_h &= \frac{1}{\sqrt{N_\mathrm{B}}} \left(\sqrt{p_h^{(d)}} \vec{a}_\mathrm{N_\mathrm{B}} (\varphi_\mathrm{BU}) + \sqrt{p_h^{(r)}} \vec{a}_\mathrm{N_\mathrm{B}} (\varphi_\mathrm{BR}) \right),\\
     \vec{f}_l &= \frac{1}{\sqrt{N_\mathrm{B}}} \left(\sqrt{p_l^{(d)}} \vec{a}_\mathrm{N_\mathrm{B}} (\varphi_\mathrm{BU}) + \sqrt{p_l^{(r)}} \vec{a}_\mathrm{N_\mathrm{B}} (\varphi_\mathrm{BR}) \right),
\end{align}
where $p_h^{(d)}$ ($p_l^{(d)}$) and $p_h^{(r)}$ ($p_l^{(r)}$) are the transmit power allocated to the HC (LC) message in the direction of the user and the RIS-direction, respectively. The phase shifts of the RIS elements are selected to maximize the received power at the user, i.e., $\mat{\Phi} = \diag\left(\vec{a}_{N_\mathrm{R}}(\varphi_\mathrm{RU}-\varphi_\mathrm{RB})\right)$. Furthermore, due to the utilization of pencil beams, we assume that $|\vec{a}_\mathrm{N_\mathrm{B}}^H (\varphi_\mathrm{BU}) \vec{a}_\mathrm{N_\mathrm{B}} (\varphi_\mathrm{BR})| \approx 0$. 
Hence, we can approximate \eqref{sinr_h} and \eqref{sinr_l} by the following expressions:
\begin{align}
\label{sinr_h_simple}
    \Gamma_h &= \frac{\beta_d N_\mathrm{B}\eta_d^2 p_h^{(d)} + \beta_r N_\mathrm{B} N_\mathrm{R}\eta_r^2 p_h^{(r)}}{\beta_d N_\mathrm{B}\eta_d^2 p_l^{(d)} + \beta_r N_\mathrm{B} N_\mathrm{R}\eta_r^2 p_l^{(r)} + \sigma_n^2},\\
\label{sinr_l_simple}
    \Gamma_l &= \frac{\beta_d N_\mathrm{B}\eta_d^2 p_l^{(d)} + \beta_r N_\mathrm{B} N_\mathrm{R}\eta_r^2 p_l^{(r)}}{\sigma_n^2}.
\end{align}
As a result, the achievable data rates are given as
\begin{align}
    R_h &= B \log_2(1 + \Gamma_h(\beta_d, \beta_r)),\\
    R_l &= B \log_2(1 + \Gamma_l(\beta_d, \beta_r)).
\end{align}

\section{Problem Formulation and Proposed Scheme}
Our goal is to serve the throughput demand given by the packet arrival rate $\bar{A}$, while providing high reliability of the data classified as HC despite the intermittent THz channel. More precisely, with the proposed superposition coding scheme, we require the HC data to be decodable via the RIS if the direct link is blocked. Hence, while the LC data stream suffers from outages whenever the direct LOS path is blocked, the HC data stream is disrupted only when the more reliable RIS-link becomes unavailable. Thereby, the outage probability of the LC data is $q_d$, whereas the HC data only exhibits an outage with probability $q_r$.

We formulate an optimization problem to find the optimal power allocation to stabilize both (HC and LC) queues for a given $\alpha$. 
Note that mean queue stability is achieved when the average service rate is greater than the arrival rate. We introduce the optimization variable $\vec{\delta} = [\delta_h, \delta_l]$ denoting the gap between the rate of successfully delivered packets and the packet arrival rate. As optimization objective, we maximize the minimum of the weighted gap variables to ensure fairness for both queues. 

Hence, defining $\vec{p} = [p_h^{(d)}, p_h^{(r)}, p_l^{(d)}, p_l^{(r)}]$ and $\vec{R} = [R_h, R_l]$, our optimization problem is formulated as follows:
\begin{subequations} \label{opt_original}
\begin{align}
\max_{\vec{\delta}, \vec{p}, \vec{R}} ~& \min\{\alpha\delta_h, (1-\alpha)\delta_l\}\tag{\theparentequation}\\
\text{s.t.}~~ &(1-q_r)\frac{T}{M} R_h - \alpha \bar{A}\geq \delta_h,  \label{HC_stability_constr}\\
 & (1-q_d) \frac{T}{M} R_l - (1-\alpha)\bar{A} \geq \delta_l,  \label{LC_stability_constr} \\
 & \delta_h, \delta_l \geq 0, \label{delta_constr} \\
&R_h \leq B \log_2\left(1+\Gamma_{h}(\beta_d, \beta_r=1)\right), \quad \beta_d\in \{0,1\}\label{HC_rate_constr}\\
&R_l \leq  B \log_2\left(1+\Gamma_{l}(\beta_d=1, \beta_r=1)\right),\label{LC_rate_constr}\\
 & p_h^{(d)} + p_h^{(r)} +p_l^{(d)} + p_l^{(r)} \leq P_\mathrm{max}. \label{power_constr}
\end{align}
\end{subequations}
Here, the constraints \eqref{HC_stability_constr}, \eqref{LC_stability_constr}, and \eqref{delta_constr} provide mean stability of the HC and LC queue. Meanwhile, \eqref{HC_rate_constr} ensures that the HC data stream can be decoded in case of an unavailable direct path, i.e., $\beta_d=0$, via the RIS-link. The LC data stream, however, is only required to be decodable if the direct link exists as given by \eqref{LC_rate_constr}. Finally, \eqref{power_constr} is the transmit power constraint of the BS. 
Note that \eqref{opt_original} is a non-convex problem due to the rate expressions. Thus, we adopt a successive convex approximation (SCA) approach to iteratively optimize the power allocation. Based on a fractional programming framework proposed in \cite{shen2018fractional}, the original problem is approximated by a convex problem via a quadratic transform that is applied to the fractional SINR expressions. The detailed problem reformulation and the iterative algorithm are given in Appendix \ref{SCA_app}.

\section{Numerical Results}
\begin{table}[tb]
    \centering
    \small{
    \begin{tabular}{|c|c|}
    \hline
      Transmit power $P_\mathrm{max}$   & 10 dBm \\ \hline
      Noise power $N_0$  &  -174 dBm/Hz\\ \hline
      Antenna gain $G_\mathrm{B}, G_\mathrm{U}$ & 20 dB\\ \hline
     Bandwidth $B$ & 10 GHz  \\ \hline
     Distances $d_\mathrm{BU}, d_\mathrm{BR}, d_\mathrm{RU}$ & 10 m, 8.7 m, 2 m  \\ \hline
     Carrier frequency $f$ & 300 GHz  \\ \hline
     Mol. absorption coeff. $k_a(f)$ & 0.0012 $\text{m}^{-1}$  \\ \hline
     BS antennas / RIS elements $N_\mathrm{B}$, $N_\mathrm{R}$ & 64, $10^4$  \\ \hline
     Blockage prob. $q_d$, $q_r$ & 0.3, 0.1 \\ \hline
    \end{tabular}}
    \caption{Simulation Parameters (if not stated otherwise).}
    \label{tab:parameters}
\end{table}
The performance of our proposed MC-SC scheme is evaluated through numerical simulations with the parameters given in Table \ref{tab:parameters}.
\begin{figure}[tb]
    \centering
    \includegraphics{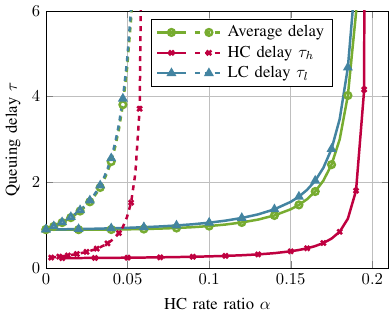}
    \caption{Average waiting time in the queue as a function of the HC rate ratio $\alpha$ for our proposed MC-SC scheme (solid lines) and an orthogonal baseline scheme (dashed lines).}
    \label{fig:delay}
\end{figure}
First, we compare the proposed scheme to an orthogonal multiple access (OMA) baseline, such as TDMA, where a fraction of each time slot is allocated to transmit the HC data stream with enhanced reliability using the RIS-path, while the remaining time resources are used to transmit LC data through the direct link. We evaluate the average queuing delay, i.e., the mean waiting time of HC/LC data in each buffer, which, according to Little's law \cite{littlesLaw}, is given as $\tau_h = \frac{\mathbb{E}\{Q_h\}}{\alpha \bar{A}}$ and $\tau_l = \frac{\mathbb{E}\{Q_l\}}{(1-\alpha) \bar{A}}$ for the HC and LC data stream, respectively. Figure \ref{fig:delay} shows the average queuing delay for a packet size of $M=10$ Mbit, time slot duration $T=100$ ms, and a total packet arrival rate of $\bar{A}=700$ packets/slot. 
With the orthogonal scheme, the delay increases as soon as a portion of the data is treated as HC data and transmitted via the RIS-link ($\alpha >0$), and the system soon becomes unstable (approx. $\alpha>0.05$). With our proposed scheme, the average queuing delay is barely affected if less than 10\% of the  packets are treated as HC data, and only a minor increase in delay is observed for up to 15\% of HC packets. The system becomes unstable approximately for $\alpha> 0.19$. 

\begin{figure}[tb]
    \centering
    \includegraphics[]{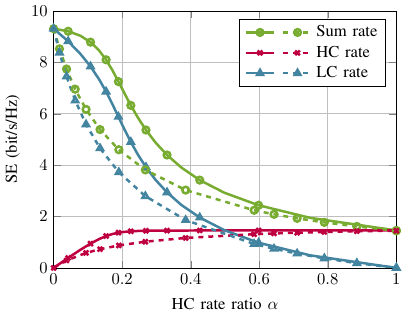}
    \caption{Average achievable SE as a function of $\alpha$ of our proposed MC-SC scheme (solid lines) compared to an orthogonal baseline scheme (dashed lines).}
    \label{fig:rate_OMA}
\end{figure}
In Figure \ref{fig:rate_OMA}, we evaluate the average achievable spectral efficiency (SE) as a function of $\alpha$ for the proposed MC-SC and the baseline scheme.
In fact, the maximum achievable sum rate is the maximum feasible $\bar{A}$ in \eqref{opt_original}, i.e., the total packet arrival rate for which marginal queue stability can be achieved ($\delta_h=\delta_l=0$). Note that this can be considered an upper bound on the feasible service rate, as the actual packet arrival rate should be smaller to prevent large queuing delays. As expected, the achievable sum rate is maximized for $\alpha=0$, i.e., when all data is transmitted through the direct LOS link. With increasing $\alpha$, the sum rate is reduced, as more power is invested for the HC data stream using the weaker RIS-path. We observe that up to a certain value of $\alpha$, the HC rate increases almost linearly, while the overall rate loss is rather small. For larger values of $\alpha$, the LC rate suffers heavily while the HC rate gradually becomes saturated. This tipping point of $\alpha$, where the rate gap between our scheme and the baseline is maximized, is at $\alpha \approx 0.15$ in Fig. \ref{fig:rate_OMA}.

Next, we analyze the impact of the blockage probability of the direct link and the number of RIS elements on the achievable rates. Recall that this work addresses the tradeoff of using the strong but highly intermittent LOS path versus exploiting the more reliable yet considerably weaker reflection path via the RIS for critical data transmission. Thus, our proposed scheme is significantly impacted by the degree of intermittency of the direct link on the one hand, and the path attenuation of the RIS-link on the other hand.

\begin{figure}[tb]
    \centering
    \includegraphics{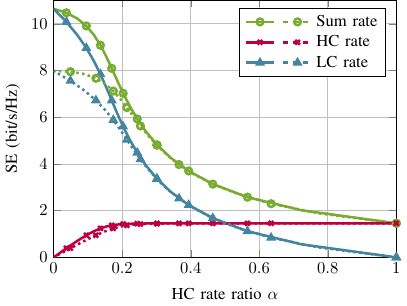}
    \caption{Average achievable SE as a function of $\alpha$ shown for different blockage probability of the direct LOS link, i.e., $q_d=0.2$ (solid) and $q_d=0.4$ (dotted).}
    \label{fig:rate_q}
\end{figure}
Figure \ref{fig:rate_q} shows the maximum average achievable SE as a function of $\alpha$ for a LOS blockage probability $q_d = 0.2$ and $q_d = 0.4$, while the blockage probability of the RIS-link remains fixed at $q_r=0.1$. We notice that while the achievable LC rate naturally becomes lower for $q_d=0.4$, the sum rate also decays slower in that case. This means that when the direct link becomes less reliable, we benefit more from transmitting a higher fraction of data packets via the RIS link, hence a larger $\alpha$ can be selected.

\begin{figure}[tb]
    \centering
    \includegraphics{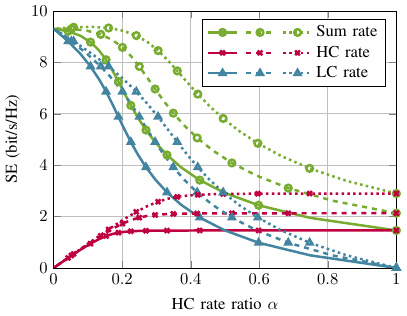}
    \caption{Average achievable SE as a function of $\alpha$ for different number of RIS elements, i.e., $N_\mathrm{R}=10^4$ (solid), $N_\mathrm{R}=2\cdot 10^4$ (dashed), and $N_\mathrm{R}=4\cdot 10^4$ (dotted).}
    \label{fig:rate_N}
\end{figure}
A similar effect is observed in Figure \ref{fig:rate_N} when, in turn, the gain of the RIS-link is increased by adding more reflect elements to the surface. Here, a larger RIS leads to a higher achievable HC rate and the sum rate is less affected by a larger value of $\alpha$. In fact, even a slight increase in the average sum rate can be observed when $N_\mathrm{R} = 4\cdot 10^4$ up to $\alpha\approx 0.15$ due to the use of the more reliable RIS-path. In general, both a stronger RIS-path and a less reliable LOS path increase the benefit of our scheme. This becomes evident as in Fig. \ref{fig:rate_q} and \ref{fig:rate_N}, we observe that the tipping point, where the HC rate starts to converge, occurs at higher values of $\alpha$ when $q_d$ or $N_\mathrm{R}$ increase. This means that a larger fraction of data can be classified as HC and transmitted with high reliability without significant loss in the overall service rate.

\section{Conclusion}
This work studies one of the key challenges of THz communication, namely dealing with the severe THz path loss on the one hand, and the sensitivity to blockage causing link intermittency on the other hand. These THz properties result in a fundamental rate-reliability tradeoff, which, in this paper, is approached from a data significance viewpoint. 
This work examines the downlink transmission of mixed-criticality data via a direct LOS link that is highly intermittent due to blockage, and a RIS-aided reflection path, which, in turn, is more reliable, yet much weaker than the direct link. A criticality-aware superposition coding based scheme is proposed, which allows for more reliable delivery of a portion of high-criticality data, whereas low-criticality data is only decoded when the LOS path is available. Adopting a data queuing model at the BS, a power allocation problem is considered to achieve queue stability. The problem is solved by an iterative SCA-based algorithm. Simulation results show that a critical data stream can be transmitted with enhanced reliability using the RIS, without compromising the queuing delay when our proposed SC scheme is used. Evaluating the maximum achievable rates for different HC/LC ratios further demonstrates the potential of our scheme, which becomes more beneficial with increasing LOS blockage probability and a sufficiently strong RIS-link.

\appendix
\subsection{SCA-based iterative optimization} \label{SCA_app}
First, we introduce auxiliary variables $\vec{\gamma}=[\gamma_h, \gamma_l]$. Hence, we can rewrite problem \eqref{opt_original} as follows:
\begin{subequations} \label{opt2}
\begin{align}
\max_{\vec{\delta}, \vec{p}, \vec{R}, \vec{\gamma}} ~& \min\{\alpha\delta_h, (1-\alpha)\delta_l\}\tag{\theparentequation}\\
\text{s.t.}~~ & \eqref{HC_stability_constr}, ~\eqref{LC_stability_constr},~ \eqref{delta_constr}, ~\eqref{power_constr}, \\
&R_h \leq B \log_2\left(1+\gamma_h\right), \label{HC_rate_constr2}\\
&R_l \leq  B \log_2\left(1+\gamma_l\right),\label{LC_rate_constr2}\\
& \gamma_h \leq \Gamma_{h}(\beta_d, \beta_r=1),\quad \beta_d\in\{0,1\},\\
& \gamma_l \leq \Gamma_{l}(\beta_d=1, \beta_r=1).
\end{align}
\end{subequations}
Next, we adopt the quadratic transform proposed in \cite{shen2018fractional} to handle the fractional SINR expressions. By introducing the auxiliary variables $\vec{\mu} = [\mu_{h,0}, \mu_{h,1}, \mu_l]$ and applying the approach from \cite{shen2018fractional}, we obtain the functions
\begin{align}
\begin{split}
g_{h,\beta_d}(\vec{p}, \vec{\gamma})& = \gamma_h - 2\mu_{h,\beta_d} 
\sqrt{\beta_d  N_\mathrm{B}N_\mathrm{R} \eta_{r}^2 p_h^{(d)} + N_\mathrm{B}N_\mathrm{R} \eta_{r}^2 p_h^{(r)}}\\ &+ {\mu_{h,\beta_d}}^2 \left(\beta_d  N_\mathrm{B}N_\mathrm{R} \eta_{r}^2 p_l^{(d)} + N_\mathrm{B}N_\mathrm{R} \eta_r^2 p_l^{(r)} + \sigma_n^2\right),
\end{split} \label{g_h}\\
\begin{split}
g_{l}(\vec{p}, \vec{\gamma})& = \gamma_l - 2\mu_l 
\sqrt{N_\mathrm{B} \eta_d^2 p_l^{(d)} + N_\mathrm{B}N_\mathrm{R} \eta_{r}^2 p_l^{(r)}} + {\mu_l}^2 \sigma_n^2.
\end{split} \label{g_l}
\end{align}
The optimal $\vec{\mu}$ for fixed $\vec{p}$ and $\vec{\gamma}$ can be obtained by setting the derivatives of \eqref{g_h} and \eqref{g_l} to zero. Thus, we have
\begin{align}
{\mu_{h,\beta_d}}^* &= \frac{\sqrt{\beta_d  N_\mathrm{B}N_\mathrm{R} \eta_{r}^2 p_h^{(d)} + N_\mathrm{B}N_\mathrm{R} \eta_{r}^2 p_h^{(r)}}}{\beta_d  N_\mathrm{B}N_\mathrm{R} \eta_{r}^2 p_l^{(d)} +N_\mathrm{B}N_\mathrm{R} \eta_r^2 p_l^{(r)} + \sigma_n^2}, \label{mu_h}\\
{\mu_l}^* &= \frac{\sqrt{N_\mathrm{B} \eta_{d}^2 p_l^{(d)} + N_\mathrm{B}N_\mathrm{R} \eta_{r}^2 p_l^{(r)}}}{\sigma_n^2}. \label{mu_l}
\end{align}
As for constant $\vec{\mu}$, the functions \eqref{g_h} and \eqref{g_l} are convex in $\vec{p}$ and $\vec{\gamma}$, the optimization problem \eqref{opt2} can be approximated by a convex problem with fixed $\vec{\mu}$:
\begin{subequations} \label{opt_convex}
\begin{align}
\max_{\vec{\delta}, \vec{p}, \vec{R}, \vec{\gamma}} ~& \min\{\alpha\delta_h, (1-\alpha)\delta_l\}\tag{\theparentequation}\\
\text{s.t.}~~ & \eqref{HC_stability_constr}, ~\eqref{LC_stability_constr},~ \eqref{delta_constr}, ~\eqref{power_constr},~ \eqref{HC_rate_constr2},~ \eqref{LC_rate_constr2}, \\
& g_{h,\beta_d}(\vec{p}, \vec{\gamma}) \leq 0, \quad \beta_d \in \{0,1\},\\
& g_l(\vec{p}, \vec{\gamma}) \leq 0.
\end{align}
\end{subequations}
Thus, the optimal power allocation is obtained via an iterative SCA-based algorithm. That is, the approximated problem \eqref{opt_convex} is solved using a convex optimization solver such as CVX \cite{cvx}, and the auxiliary variables $\vec{\mu}$ are alternately updated following \eqref{mu_h} and \eqref{mu_l}. The algorithm is summarized in Alg. \ref{alg:MCSC_alg}.
\begin{algorithm}[]
\caption{Power Allocation for the MC-SC scheme}
  \begin{algorithmic}[1]
    \State Initialize $\vec{p}$
    \Repeat 
        \State Compute $\mu_h$ and $\mu_l$ based on \eqref{mu_h},  \eqref{mu_l}
        \State Solve \eqref{opt_convex} for fixed $\vec{\mu}$
    \Until{Convergence}
    \end{algorithmic}
    \label{alg:MCSC_alg}
\end{algorithm}
\vspace{-2mm}

\bibliographystyle{./bibliography/IEEEtran}
\bibliography{./bibliography/IEEEabrv,./bibliography/IEEEexample,./bibliography/references}

\begin{thebibliography}{10}
\providecommand{\url}[1]{#1}
\csname url@samestyle\endcsname
\providecommand{\newblock}{\relax}
\providecommand{\bibinfo}[2]{#2}
\providecommand{\BIBentrySTDinterwordspacing}{\spaceskip=0pt\relax}
\providecommand{\BIBentryALTinterwordstretchfactor}{4}
\providecommand{\BIBentryALTinterwordspacing}{\spaceskip=\fontdimen2\font plus
\BIBentryALTinterwordstretchfactor\fontdimen3\font minus \fontdimen4\font\relax}
\providecommand{\BIBforeignlanguage}[2]{{%
\expandafter\ifx\csname l@#1\endcsname\relax
\typeout{** WARNING: IEEEtran.bst: No hyphenation pattern has been}%
\typeout{** loaded for the language `#1'. Using the pattern for}%
\typeout{** the default language instead.}%
\else
\language=\csname l@#1\endcsname
\fi
#2}}
\providecommand{\BIBdecl}{\relax}
\BIBdecl

\bibitem{saad2021sevenTHz}
C.~Chaccour, M.~N. Soorki, W.~Saad, M.~Bennis, P.~Popovski, and M.~Debbah, ``{Seven defining features of terahertz (THz) wireless systems: A fellowship of communication and sensing},'' \emph{IEEE Commun. Surveys Tuts.}, vol.~24, no.~2, pp. 967--993, Jan. 2022.

\bibitem{akyildiz2018combating}
I.~F. Akyildiz, C.~Han, and S.~Nie, ``{Combating the distance problem in the millimeter wave and terahertz frequency bands},'' \emph{IEEE Commun. Mag.}, vol.~56, no.~6, pp. 102--108, June 2018.

\bibitem{kokkoniemi2019NLOS300}
J.~Kokkoniemi, J.~Lehtom\"aki, and M.~Juntti, ``{LOS and NLOS Channel Models for Indoor 300 GHz Communications},'' in \emph{16th Int. Symp. Wireless Commun. Syst. (ISWCS)}, 2019, pp. 441--445.

\bibitem{karacoraTHzTCOM}
Y.~Karacora, C.~Chaccour, A.~Sezgin, and W.~Saad, ``{Event-Based Beam Tracking With Dynamic Beamwidth Adaptation in Terahertz (THz) Communications},'' \emph{IEEE Trans. Commun.}, vol.~71, no.~10, pp. 6195--6210, 2023.

\bibitem{aman2023downlink}
W.~Aman, N.~Kouzayha, M.~M.~U. Rahman, and T.~Y. Al-Naffouri, ``{On the Downlink Coverage Performance of {RIS}-Assisted {THz} Networks},'' \emph{arXiv preprint arXiv:2302.02465}, 2023.

\bibitem{chaccour2020risk}
C.~Chaccour, M.~N. Soorki, W.~Saad, M.~Bennis, and P.~Popovski, ``{Risk-based optimization of virtual reality over terahertz reconfigurable intelligent surfaces},'' in \emph{Proc. IEEE Int. Conf. Commun. (ICC)}, Dublin, Ireland, June 2020, pp. 1--6.

\bibitem{zarini2023RISaided}
H.~Zarini, N.~Gholipoor, M.~R. Mili, M.~Rasti, H.~Tabassum, and E.~Hossain, ``{Resource Management for Multiplexing {eMBB} and {URLLC} Services Over {RIS}-Aided THz Communication},'' \emph{IEEE Trans. Commun.}, vol.~71, no.~2, pp. 1207--1225, 2023.

\bibitem{chaccour2020HRLLC}
C.~Chaccour, M.~N. Soorki, W.~Saad, M.~Bennis, and P.~Popovski, ``{Can terahertz provide high-rate reliable low latency communications for wireless VR?}'' \emph{IEEE Internet Things J.}, Jan. 2022.

\bibitem{boulogeorgos2021directional}
A.-A.~A. Boulogeorgos, J.~M. Jornet, and A.~Alexiou, ``{Directional Terahertz Communication Systems for 6G: Fact Check},'' \emph{IEEE Veh. Technol. Magazine}, vol.~16, no.~4, pp. 68--77, 2021.

\bibitem{sarieddeen2021terahertz}
H.~Sarieddeen, A.~Abdallah, M.~M. Mansour, M.-S. Alouini, and T.~Y. Al-Naffouri, ``{Terahertz-band {MIMO-NOMA}: Adaptive superposition coding and subspace detection},'' \emph{IEEE Open J. Commun. Soc.}, vol.~2, pp. 2628--2644, 2021.

\bibitem{choi2015multicast}
J.~Choi, ``{Minimum Power Multicast Beamforming With Superposition Coding for Multiresolution Broadcast and Application to {NOMA} Systems},'' \emph{IEEE Trans. Commun.}, vol.~63, no.~3, pp. 791--800, 2015.

\bibitem{reifert2023comeback}
R.-J. Reifert, S.~Roth, A.~A. Ahmad, and A.~Sezgin, ``{Comeback Kid: Resilience for Mixed-Critical Wireless Network Resource Management},'' \emph{IEEE Transactions on Vehicular Technology}, pp. 1--17, 2023.

\bibitem{karacora2023rate}
Y.~Karacora and A.~Sezgin, ``{Rate-splitting enabled multi-connectivity in mixed-criticality systems},'' in \emph{Proc. IEEE Int. Conf. Commun. (ICC)}, Rome, Italy, June 2023, pp. 5340--5345.

\bibitem{shen2018fractional}
K.~Shen and W.~Yu, ``{Fractional programming for communication systems--Part I: Power control and beamforming},'' \emph{IEEE Trans. Signal Process.}, vol.~66, no.~10, pp. 2616--2630, 2018.

\bibitem{sarieddeen2021overview}
H.~Sarieddeen, M.-S. Alouini, and T.~Y. Al-Naffouri, ``{An overview of signal processing techniques for terahertz communications},'' \emph{Proc. IEEE}, Oct. 2021.

\bibitem{kokkoniemi2021line}
J.~Kokkoniemi, J.~Lehtom{\"a}ki, and M.~Juntti, ``{A line-of-sight channel model for the 100--450 gigahertz frequency band},'' \emph{EURASIP J. Wireless Commun. Netw.}, vol. 2021, no.~1, pp. 1--15, Apr. 2021.

\bibitem{littlesLaw}
J.~D.~C. Little, ``{A Proof for the Queuing Formula: {$L= \lambda W$}},'' \emph{Operations Research}, vol.~9, no.~3, pp. 383--387, 1961.

\bibitem{cvx}
M.~Grant and S.~Boyd, ``{CVX}: Matlab software for disciplined convex programming, version 2.1,'' \url{http://cvxr.com/cvx}, Mar. 2014.

\end{thebibliography}

\end{document}